\documentclass[12pt,preprint]{aastex}

\begin{document}

\newcommand{\lya}{Lyman-$\alpha$}
\newcommand{\eqw}{\hbox{EW}}
\newcommand{\ifilt}{$i_{775}$}
\newcommand{\zfilt}{$z_{850}$}
\newcommand{\Vfilt}{$V_{606}$}
\newcommand{\Bfilt}{$B_{450}$}
\newcommand{\Jfilt}{$J_{110}$}
\newcommand{\Hfilt}{$H_{160}$}
\def\erg{\hbox{erg}}
\def\cm{\hbox{cm}}
\def\sec{\hbox{s}}
\def\f17{f_{17}}
\def\Mpc{\hbox{Mpc}}
\def\Gpc{\hbox{Gpc}}
\def\nm{\hbox{nm}}
\def\km{\hbox{km}}
\def\kms{\hbox{km s$^{-1}$}}
\def\year{\hbox{yr}}
\def\Myr{\hbox{Myr}}
\def\deg{\hbox{deg}}
\def\arcsec{\hbox{arcsec}}
\def\microJy{\mu\hbox{Jy}}
\def\zre{z_r}
\def\fesc{f_{\rm esc}}
\def\lstar{\ifmmode {L_\star}\else
                ${L_\star}$\fi}
\def\mstar{\ifmmode {m_\star}\else
                ${m_\star}$\fi}
\def\phistar{\ifmmode {\phi_\star}\else
                ${\phi_\star}$\fi}

\def\ergcm2s{\ifmmode {\rm\,erg\,cm^{-2}\,s^{-1}}\else
                ${\rm\,ergs\,cm^{-2}\,s^{-1}}$\fi}
\def\ergsec{\ifmmode {\rm\,erg\,s^{-1}}\else
                ${\rm\,ergs\,s^{-1}}$\fi}
\def\kmsMpc{\ifmmode {\rm\,km\,s^{-1}\,Mpc^{-1}}\else
                ${\rm\,km\,s^{-1}\,Mpc^{-1}}$\fi}
\def\kpc{{\rm kpc}}
\def\nv{\ion{N}{5} $\lambda$1240}
\def\civ{\ion{C}{4} $\lambda$1549}
\def\oii{[\ion{O}{2}] $\lambda$3727}
\def\oiipair{[\ion{O}{2}] $\lambda \lambda$3726,3729}
\def\oiii{[\ion{O}{3}] $\lambda$5007}
\def\oiiipair{[\ion{O}{3}] $\lambda \lambda$4959,5007}
\def\taulya{\tau_{Ly\alpha}}
\def\llya{L_{Ly\alpha}}
\def\nbar{\bar{n}}
\def\Msun{M_\odot}
\def\msun{M_\odot}

\title{An overdensity of galaxies at $z=5.9 \pm 0.2$ in the Ultra Deep Field confirmed using the ACS grism}

\author{
S. Malhotra\altaffilmark{1},
J.E. Rhoads\altaffilmark{1},
N. Pirzkal\altaffilmark{1},
Z. Haiman\altaffilmark{2},
C. Xu\altaffilmark{1},
E. Daddi\altaffilmark{3,4,5},
H. Yan \altaffilmark{6},
L.E. Bergeron\altaffilmark{1},
J. Wang\altaffilmark{11} ,
H.C. Ferguson \altaffilmark{1},
C. Gronwall \altaffilmark{9},
A. Koekemoer\altaffilmark{1},
M. Kuemmel\altaffilmark{4},
L. A. Moustakas\altaffilmark{1},
N. Panagia\altaffilmark{1,10},
A. Pasquali \altaffilmark{8},
S. di Serego Alighieri\altaffilmark{12},
M. Stiavelli\altaffilmark{1},
J. Walsh\altaffilmark{4},
R. A. Windhorst\altaffilmark{7}}

\begin{abstract}

We present grism spectra taken with the Advanced Camera for Surveys to
identify 29 red sources with (\ifilt-\zfilt) $ \ge 0.9$ in the Hubble
Ultra Deep Field (HUDF). Of these 23 are found to be galaxies at
redshifts between z=5.4 and 6.7, identified by the break at 1216 \AA\
due to IGM absorption; two are late type dwarf stars with red colors;
and four are galaxies with colors and spectral shape similar to dust
reddened or old galaxies at redshifts z $\approx 1-2$. This
constitutes the largest uniform, flux-limited sample of
spectroscopically confirmed galaxies at such faint fluxes (\zfilt
$\le$ 27.5).  Many are also among the most distant spectroscopically
confirmed galaxies (at redshifts up to z=6.7).

We find a significant overdensity of galaxies at redshifts $z=5.9 \pm
0.2$. Nearly two thirds of the galaxies in our sample (15/23) belong
to this peak. Taking into account the selection function and the
redshift sensitivity of the survey, we get a
conservative overdensity of at least a factor of two along the
line-of-sight. The galaxies found in this redshift peak are also
localized in the plane of the sky in a non-random manner, occupying about
half of the ACS chip. Thus the volume overdensity is a factor of
four. The star-formation rate derived from detected sources 
in this overdense region is sufficient to reionize the local IGM.

\end{abstract}

\altaffiltext{1}{Space Telescope Science Institute, 3700 San Martin Drive, Baltimore, MD21218, USA}
\altaffiltext{2}{Department of Astronomy, Columbia University, 1328 Pupin Hall, 550 West 120th St., NY, 10027, USA}
\altaffiltext{3}{ESO/ST-ECF, Karl-Schwarschild-Strasse 2, D-85748, Garching bei M\"{u}nchen, Germany}
\altaffiltext{4}{National Optical Astronomy Observatories, Cherry Avenue, Tucson, AZ, , USA}
\altaffiltext{5}{Spitzer Fellow}
\altaffiltext{6}{Spitzer Science Center, California Institute of Technology, Mail-Stop 100-22, Pasadena, CA 91125}
\altaffiltext{7}{Dept. of Physics \& Astronomy, Arizona State University, P.O. Box 871504, Tempe, AZ 85287-1504, USA}
\altaffiltext{8}{Institute of Astronomy, ETH H\"{o}nggerberg, 8093 Zurich, Switzerland}
\altaffiltext{9}{Dept of Astronomy, Pennsylvania State University, 525 Davey Laboratory, University Park, PA 16802}
\altaffiltext{10}{On assigment from Space Telescope Operations Division, Research and Scientific Support Department of ESA}
\altaffiltext{11}{Center for Astrophysics, University of Science and Technology of China, Hefei, Anhui 230026, P. R. China}
\altaffiltext{12}{INAF- Osservatorio Astrofisico di Arcetri, Largo E. Fermi, 5. I-50125 Firenze, Italia}


\section{Introduction}

While theory (e.g. Press \& Schechter 1974) and numerical simulations can
tell us about the gravitational collapse and clustering of dark matter, the 
onset of star-formation in galaxies is complicated enough that observations
are required to guide the theory. The question about how biased the early
star-formation is can be addressed by measuring clustering at the highest 
redshifts accesible.

The Hubble Ultra Deep Field provides a uniquely deep look at the
universe in its infancy. With visible and near-infrared observations
reaching a depth of \zfilt $ =28.2 $ and $ J =26.85 $ AB magnitudes
(10$\sigma$ for $0."5$ aperture) (Beckwith et al. 2004 in prep.,
Thompson et al. 2005, Bouwens et al. 2004), one can reliably detect
galaxies as faint as $0.02 \times L_{*}$ at $z \approx 6$ (Yan \&
Windhorst 2004, Bunker et al. 2004). Because of its depth this is
potentially a good field to study formation and clustering of
galaxies, the only handicap being its small size (1.26 physical Mpc on
a side at z$\approx 6$).

The census of galaxies at z$\approx 6$ is also interesting because the
Gunn-Peterson trough has been observed near this redshift (Fan et al
2002), suggesting that reionization ended around z=6.  On the other
hand, evidence from microwave background observations suggests that
substantial reionization likely occurred at $z\sim 15$ (Spergel et
al. 2003), and \lya\ emitters (Rhoads et al. 2004, Hu et al. 2002, and
Taniguchi et al. 2004) suggest that reionization was largely complete
by redshift z $\approx 6.5$ (Malhotra \& Rhoads 2004, hereafter
MR04). Strong clustering of galaxies at this epoch would complicate
the reionization scenarios, leading to inhomogeneous reionization.

In accounting for the source(s) of reionizing photons at the epoch of
reionization, quasars and active galactic nuclei are not sufficient
(Barger et al. 2003, Moustakas \& Immler 2004, Wang et al. 2004) and
even a hitherto undetected population of faint or obscured quasars
would not be able to reionize the IGM at z~6 without overproducing the
unresolved soft X-ray background (Dijkstra etal)" . Therefore galaxies
must provide a large part of the ionizing photon budget. One of the
major goals of the very faint imaging in the Hubble Ultra Deep Field
is to determine the luminosity function and therefore the ionizing
photon budget of the galaxies at z $\approx$ 6. Bunker et al. 2004,
Yan et al. 2004, Stiavelli et al. 2004 have all carried out the
determination of the luminosity function at $z \approx 6$ on the basis
of imaging data alone, using a color selection $(i-z) > 1.3$ to select
high redshift galaxies.

We have carried out deep unbiased spectroscopy in the Hubble Ultra
Deep Field with the ACS grism to spectroscopically confirm these
sources, with GRAPES (GRism ACS Program for Extragalactic
Science). About 40 orbits (100 Ks) of exposure time went into the
spectroscopic follow-up.  Details of the observations and data
reduction are described in a paper by Pirzkal et al. (2004.)  The
benefits of spectroscopy are several. First, we are able to confirm a
substantial fraction, but not all, of (\ifilt - \zfilt) dropouts as
high redshift galaxies. Second, the spectra give the slope of the
continuum emission, which can constrain stellar populations (modulo
extinction). The third advantage of using the spectra to select
objects is the easily characterized selection function. The initial
color selection can be very inclusive and thus complete, if spectra
are finally used to identify objects. Finally, we get redshifts
accurate to $\Delta z \la 0.15$, which is essential for studying
clustering.

In section 2, we describe the selection of candidates and spectral
confirmation. Section 3 describes a comparison of redshifts of objects
with their expected and observed broad-band colors. In section 4 we
explore the overdensity found at z=5.9, and section 5 contains
discussion and conclusions. Appendix A contains the grism spectra of
the 23 high redshift galaxies and 4 intermediate redshift red galaxies
along with cutouts from images in the \Bfilt, \Vfilt, \ifilt, \zfilt, 
\Jfilt, and \Hfilt\ filters.

\section{Spectroscopic confirmation}

\subsection{Candidate Selection}

For candidate selection we used the source catalog released by the UDF
team, which is based on the i-band image, supplemented by sources
which were detected in the z-band only. For the very red sources
(\ifilt-\zfilt)$ > 0.9$ magnitudes we use the z-band detection parameters (e.g.
size, RA, DEC) because the signal-to-noise ratio is higher in the
z-band for such red sources. 

High redshift candidates were selected by their red
color in the ACS F775W and F850LP filters (which correspond to
SDSS $i\prime$ and $z\prime$ filters; see Beckwith et al. 2004).
The selection criteria were \\ 
(1) Red color in the \ifilt, \zfilt bands: (\ifilt-\zfilt)$ > 0.9$ magnitudes;
 which should pick up sources with redshift $z > 5.4$\\
(2) no detected flux in the F435W (``B'') filter ($B > 28.7$); and \\
(3) \zfilt $< 28.5$ magnitude.  Although our spectra typically reach depth
$z' \approx 27.1$, fainter broad band \zfilt magnitudes may be achieved
when either (a) the redshift $z > 6$, so that part of \zfilt bandpass
contributes only noise (and no signal) to the measurements, or
(b) the galaxy has a prominent emission line.

The usual criteria for selecting Lyman-break galaxies (LBGs) are oriented
towards minimizing the interlopers. One either takes a stringent color
cut, so that one has few interlopers (e.g. red galaxies at
intermediate redshifts) - both Bunker et al. 2004 and Yan \& Windhorst
2004 take a color cut of (\ifilt-\zfilt)$> 1.3$. The other approach is to
demand a red color across the break and a blue color longward of the
break (e.g.  Steidel et al. 1995, Giavalisco 2002). For $z\sim 6$
galaxies in the HUDF, this approach requires near-infrared data.
The only NIR images of the UDF that reach approximately sufficient
depth (Bouwens et al 2004) cover only a fraction of the full UDF.
With slitless spectroscopy we are able to obtain
spectra of all the objects, so we choose to be generous in our
color-cut to define the sample in order to improve completeness.

With these cuts we obtained 106 candidates.  Among these, 25 are brighter
than \zfilt(AB)=27.1 and 45 are brighter than \zfilt=27.5.  The nominal
sensitivity limit for the GRAPES spectra is between \zfilt=27.1 and 27.5,
depending on the redshift. (A fixed continuum flux density at
$1300(1+z)$\AA\ corresponds to a fainter magnitude in \zfilt-band for
higher redshift objects, where intergalactic absorption removes most
flux from the blue end of the filter bandpass.)  Comparison of
candidates so selected with the catalogs of Yan \& Windhorst 2004 and
Bunker et al. 2004, shows good agreement, missing only one object from 
the YW04 sample which lies close to another galaxy. It has been
difficult to extract the spectra of some of the objects where high
redshift candidates are close to other objects (see the next
section). Since this is a random occurance it does not introduce a bias
into our sample of spectroscopically selected galaxies.

Seven of these sources with \zfilt $<27.5$ have spectra that overlap
substantially with brighter sources nearby, and will not be considered
any further. Another four lie outside the GRAPES region, which has
about 85\% overlap with the UDF region. Out of the remaining 34, 29
have useful spectra, the rest have low s/n spectra because of a
combination of low surface brightness and rejection of a significant
fraction of the data by contamination. Thus an incompleteness
correction of about a factor of (46/29) or about 1.6 should be applied
to any results derived from the spectroscopic sample alone, since we were
not able to determine the nature of some sources due to spectral overlap.

\subsection{Grism spectra} 

About 100 kiloseconds of grism exposures can potentially provide
spectra of {\it all} sources in an unbiased way. In practice some
information is lost because of overlap of the spectra. The spectra are
subject to more crowding than the images since for each object the
light is dispersed over 100 pixels rather than a few in one
dimension. To mitigate the overlap of the spectra the grism data was
taken at four roll angles with position angles of 126, 134, 217, and
231 degrees. Previous grism observations of a supernova field yielded
another 24 ksec of data at a PA of 117 degrees.  In standard aXe
reductions (Pirzkal et al. 2001), the parts of the spectra that have
overlap with others are flagged as contaminated.  In our analysis we
have modified the code to flag contamination not just as a yes/no
binary decision but as an estimate of how much of the flux comes from
the contaminant (see Pirzkal et al. 2004 for more details). So a
spectrum of a bright source contaminated by a faint source is still
usable. In the present paper we reject all pixels contaminated by
light that is estimated to be more than 33\% of the source light. As a
further check on the contamination by other spectra, we demand that
the broad-band flux from imaging agree with the sum over that passband
in grism data.

\subsection{Interlopers}

Two sources are identified as dwarf stars: UDF~443 \footnote{The 
object numbers in this paper follow the catalog numbers from the 
officially released i-band catalog h\_udf\_wfc\_V1\_i\_cat.txt} is 
an L dwarf and UDF~366 is an M dwarf, on the basis of their spectra
 and compact spatial profiles (see Pirzkal et al. 2004b for a complete list of
unresolved sources in the HUDF). Similarly four objects--- UDF8038,
UDF8238, UDF6676 and UDF3551--- are definitely identified as red
galaxies at moderate redshifts based on the absence of a spectral break
in the grism spectra and near infrared colors (Figure \ref{spectra-red}).
Daddi et al. 2005, identify UDF8238 as an intermediate redshift (z=1.39)
galaxy along with some other red galaxies on the basis of GRAPES spectra.

\subsection{High redshift galaxies}

Twenty three galaxies are identified as high redshift galaxies on the
basis of their spectra. This identification is based on detecting the
Lyman break in the continuum for 22 of these sources and \lya\ line
and break in one source. For sources as bright as UDF2225, identifying
a Lyman break is unambiguous (see Figure \ref{spectra} for this spectrum). For
most other galaxies the signal-to-noise ratio is low, and thus the
following procedure was adopted to select reasonable spectral
confirmations. The spectra were fit with a Lyman break galaxy template
with a power law spectrum of slope $\alpha=0.2$ (where $f_{\nu}\propto
\nu^{\alpha}$) attenuated by IGM absorption calculated according to
the prescription of Madau 1995.  We do a grid search on the parameters
redshift and flux at 1250 \AA\ to determine the best fit.  With the
low s/n, the $\chi^2$ per degree of freedom is generally less than 1
\footnote{This makes us suspect that the error-bars on the grism
fluxes are too large. See Pirzkal et al. (2004) for details}. For the
sources to be identified as high redshift galaxies we require that (A)
the chi-square per degree of freedom be about one; (B) the
combined flux redward of the break is well detected, with net $s/n \ge
3$; and (C) the broad-band fluxes are consistent with the grism flux
and the fitted LBG spectrum, even though we do not fit to broad-band
fluxes while determining redshifts. With regard to point (C) we
tolerate about 30\% discrepancy in the flux between broad-band and
grism measurements, because the aperture match between imaging and the
grism often results in discrepancies of that order.

\section{Comparison with color selection}

\begin{figure}[htb]
\epsscale{0.5}
\plotone{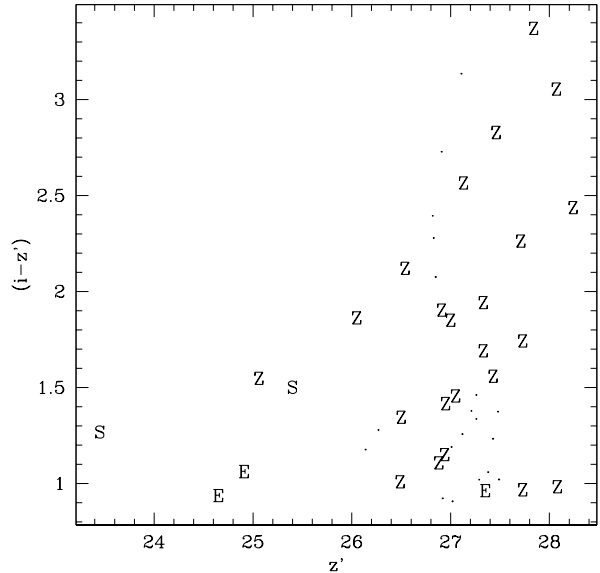}
\caption{ The identity of complete color selected sample
 in the Hubble Ultra Deep Field is shown in this color magnitude plot, the axes
of which span the color and magnitude range considered here.
 labels 'Z' indicate high redshift 
galaxy, 'S' a star, and E indicates an ERO type of galaxy. Small dots denotes
galaxies for which we did not have enough data to identify them, either 
because the galaxies fell outside the GRAPES field, or because of overlap
and contamination by other spectra.
\label{id2}}
\end{figure}

Among the 29 objects that we have examined to have clean
(non-contaminated), relatively high S/N spectra, 23 are high redshift
galaxies.  This implies about 80\% chance of finding a high redshift
galaxy with a color selection of (\ifilt-\zfilt) $> 0.9$. The five
galaxies with spectral shapes indicating old or reddened populations
at medium redshifts have colors of (\ifilt-\zfilt)$\approx 1.0$. Thus
a color cut of (\ifilt-\zfilt) $\approx 1.3$ adopted by Yan and
Windhorst 2004 and by Bunker et al. 2004 is succesful at eliminating
moderate redshift galaxies. The M-dwarf stars identified
spectroscopically have (\ifilt-\zfilt)  colors similar to the high redshift
galaxies, but are not as faint (Pirzkal et al. 2005). Figure \ref{id2} shows
the color magnitude distribution of the spectroscopically confirmed high
 redshift galaxies, stars, and moderate redshift red galaxies.

In Figure \ref{col_z} we plot the (\ifilt-\zfilt) color of the
spectroscopically confirmed galaxies against the redshift found by
fitting a Lyman-break galaxy template. Superposed on that distribution
is the expected color for star-forming galaxies from Bruzual \&
Charlot 2003 models with stellar population ages of 1,10,100 and 1000
Myrs assuming constant star-formation models with solar metallicity.
We see that the observed color-redshift distribution follows the
theoretical curves reasonably. This is not suprising, since the main
effect determining the (\ifilt-\zfilt) color is the shifting of the
Lyman break to longer wavelengths with redshift. There are color
variations in the galaxies and some are bluer or redder than the
models.  The scatter in colors seen is larger than the models would
indicate, but are not more deviant than $3 \sigma$ in the color
errors.  The redder than expected color can be explained by invoking
dust. The bluer color could be due to the presence of \lya\ line
emission which is not distinctly seen due to the low resolution of the
spectra, but can make
 (\ifilt - \zfilt) bluer for galaxies upto $ z
= 6$. Some of the apparent color scatter may be due to systematic
errors in photometry due to the presence of neighbours. There is also
an intrinsic color variation seen in two pairs of sources (3377/3398 and
3317/3325). Both pairs are resolved into separate objects in the
official HUDF catalog but  have the same redshift based on the
break seen in the spectra. The difference in the (\ifilt - \zfilt)
color is 0.7 magnitudes for the 3317/3325 pair and 0.4 magnitudes for
the 3377/3398 pair. This shows that there is likely to be a fair
variation of (\ifilt - \zfilt) colors within the same object.
Whatever the reason for the color scatter, Figure \ref{col_z} shows
that it would lead to 20-30\% incompleteness with the
(\ifilt-\zfilt)$> 1.3$ cut for $z \approx 6$ galaxies (YW04, Bunker et
al. 2004).

\begin{figure}[htb]
\epsscale{0.5}
\plotone{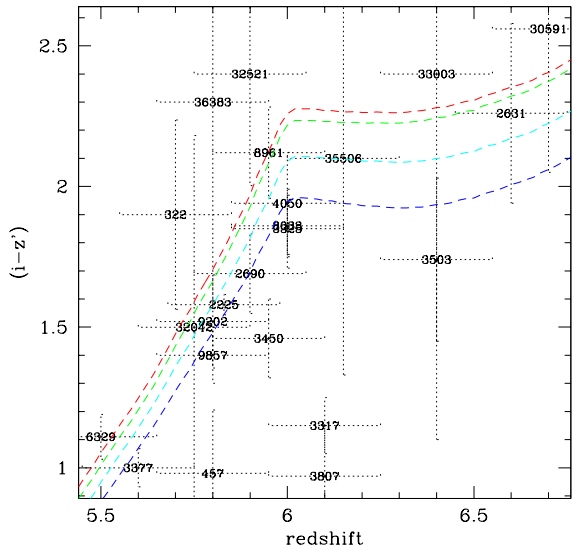}
\caption{ The (\ifilt-\zfilt) colors of galaxies are plotted against the 
redshift determined from grism spectra. The agreement with the theoretical
 curves  is generally good, although we see a bias towards slightly blue 
colors, that might be due to the presence of \lya\ line.The curves show  
colors of galaxies vs redshift using Bruzual \& Charlot models of ages $10^6$,
$10^7$,$10^8$,$10^9$ years going from bottom to top (or blue to red).
\label{col_z}}
\end{figure}

\section{Overdensity at $z=5.9 \pm 0.2$}

Figure \ref{hist_z} shows the redshift distribution of the spectroscopically
confirmed galaxies. We see an overdensity at $z=5.9 \pm 0.2$. An
overdensity was suggested by Stanway et al. (2004) on the basis of
three galaxies in and near the UDF. We see it confirmed here, with 15
galaxies in the redshift range instead of 6 expected from the lower
and higher redshifts in this sample. Thus the overdensity is 3.5
$\sigma$, assuming Poisson statistics. Doing the most na\"{\i}ve and
straightforward estimate, we find that about 2/3rd of the galaxies in
the sample are in 1/3rd of the volume, implying an overdensity of a
factor of four.

\begin{figure}[h]
\epsscale{0.5}
\vspace{2in}
\plotone{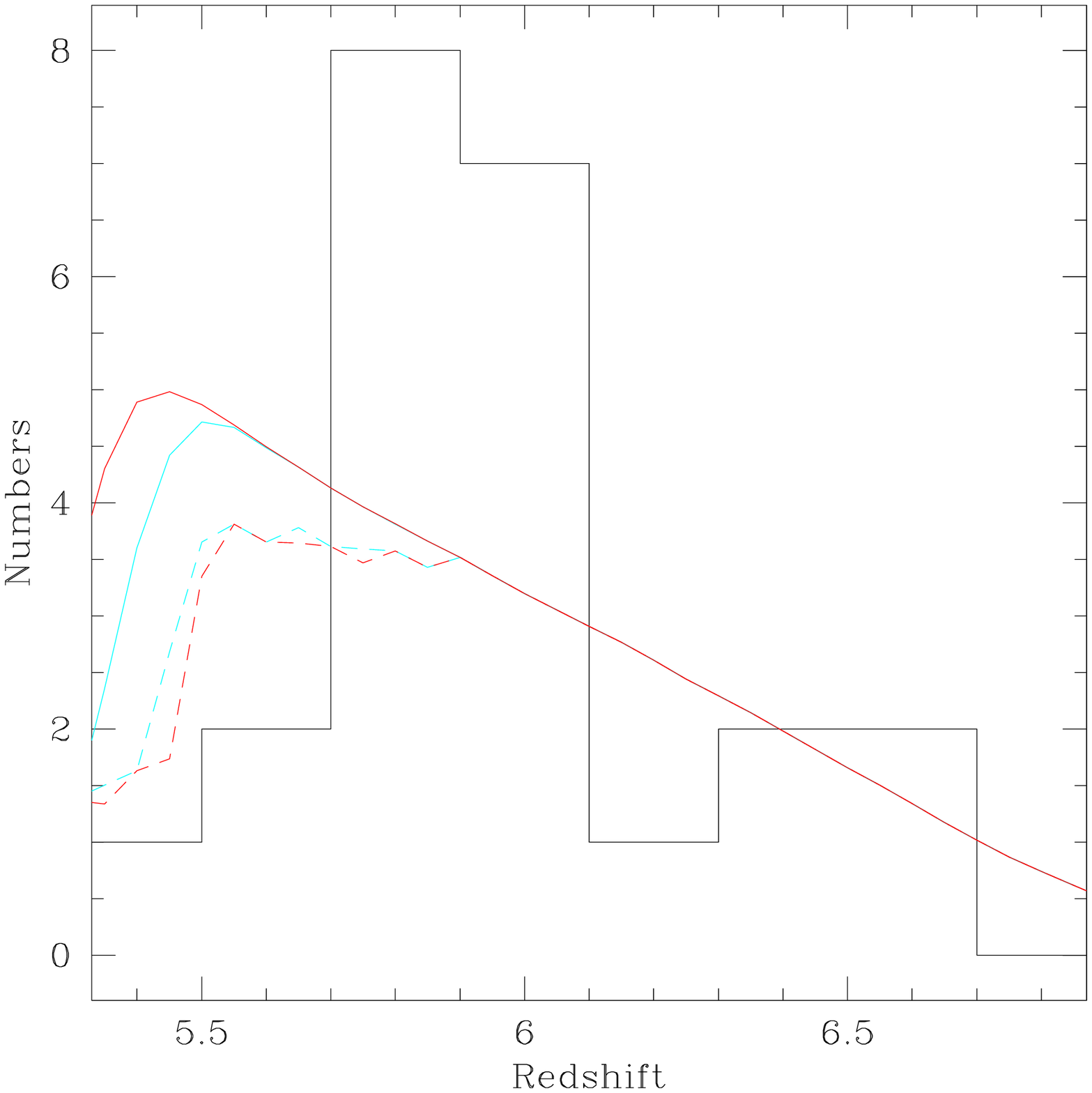}
\caption{ The redshift histogram of galaxies. We see a definite overdensity 
at $z=5.9 \pm 0.2$. Superposed on the histogram are curves of expected numbers
of galaxies in these bins. Toward the high redshift bins the expected numbers
fall off due to lack in sensitivity. In the low redshift regime the expected 
number of objects drops because the color selection misses the bluer objects. 
The solid lines show the expected numbers of objects with spectra like the
Bruzual and Charlot 2003 models with reddening of E(B-V)$= 0.15 \pm 0.15$,
for a range of 1(lower, blue curve) and 1000 (upper, red curve) Myrs old 
stellar populations. The dashed lines show the expected numbers if the 
objects have the color range seen at $5.7 \le z \le 6.1$ in our sample.}
\label{hist_z}
\end{figure}

This overdensity cannot be explained by selection effects alone. In
Figure \ref{hist_z} we over-plot the expected number density of
objects.  Folding in the color selection, the grism response function
and a Lyman break galaxy luminosity function from YW04, we predict the
number of galaxies in each redshift bin, shown in figure
\ref{hist_z}. The drop-off at the higher redshift is due to a decline
in the sensitivity of the grism at the red end, the roll-off at the
low redshift in the models is due to the color selection, which starts
to make us lose the bluer end of the z=5.5 sample. The solid curves
show selection function with the color variation introduced by
supposing a reddening with mean E(B-V)=0.15 magnitudes and
$\sigma(E_{(B-V)}) =$ 0.15 magnitudes, which is standard for Lyman
break galaxies (Giavalisco, private comm.); the underlying stellar
populations are 1 and 1000 Myrs 
old for the two curves. The dashed curves show the selection function
at z=5.5 if we use the wider color range empirically seen in our
sample at z=5.7-6.1. In the end the total number of i-drops matches
the YW04 number and the overdensity at z=5.9 is seen to be a factor of
2. Thus a factor of two is the conservative lower bound on the peak at
overdensity at $z\approx 6$, averaged over the field of view.

The true overdensity could be higher if the peak is substantially
spread out by low redshift resolution in the grism data.  With the low
resolution spectra afforded by the ACS Grism, we are able to determine
the redshifts to an accuracy of $\Delta z\approx 0.15$ for a typical
object of size 0."25.  Besides this, \lya\ emission can bias the
redshift measurements obtained from the Lyman break, even if the line
is too weak to appear obvious in the low-resolution grism spectrum.
The shift in the estimate is described by $ (\exp{(- EW / \sigma)} -
1) \times \sigma $ where $2.355 \sigma$ is the FWHM resolution of the
spectrum.  For typical values in the GRAPES survey, this corresponds
to a shift of $\la 200$\AA, or a redshift change of $\la 0.15$. This
implies that $\Delta z=0.15-0.2$. So any spike in the distribution of
galaxies is smeared by up to $\Delta z=0.2$ due to the limited
wavelength resolution of the grism.

The overdensity at z=5.9 is also supported by complementary data on
\lya\ emitters from CTIO by Wang, Malhotra \& Rhoads (2005) who have
imaged a large area, 36' $=12.9$ Mpc on a side, including the Hubble
Ultra Deep Field. They show that the HUDF sits on the edge of a much
larger scale ($> 3 $ Mpc) structure traced by \lya\ emitters at
z=5.7-5.77. Wang et al. 2005 report roughly a factor 3-4 overdensity
in the \lya\ emitters at z $\approx 5.8$, compared to other studies of
\lya\ emitters at z=5.8 (Rhoads et al. 2003, Rhoads \& Malhotra 2001,
Hu et al.  2003, MR04).

Figure \ref{lss} shows the distribution in the sky of the 15 galaxies
in the redshift range $z = 5.9 \pm 0.2$ in the HUDF along with the
larger scale structure seen in the \lya\ emitters at z=5.7-5.77.  We
see that the galaxy distribution in the HUDF continues the voids seen
in the larger distribution. Even within the HUDF the $z\sim 6$
galaxies are not distributed uniformly, and avoid one corner of the
field. We applied a two dimensional Kolmogrov-Smirnov test (Peacock
1983, Fasano \& Franceschini 1987, Press et al. 1992) to the
distribution of the 15 galaxies at $z = 5.9 \pm 0.2$. Even with just
15 galaxies, the test gives only a 5\% chance that they could be so
arranged due to chance.  Large scale structure has been seen in
several such studies for both Lyman Break galaxies and \lya\ emitters
(e.g., Steidel et al. 1998, 2000; Venemans et al. 2002; Miley et al. 
2004; Palunas et al. 2004; Ouchi et al. 2001, 2003; Shimasaku et al. 
2003; Foucaud et al. 2003; and references therein).

\begin{figure}[ht]
\epsscale{0.9}
\plottwo{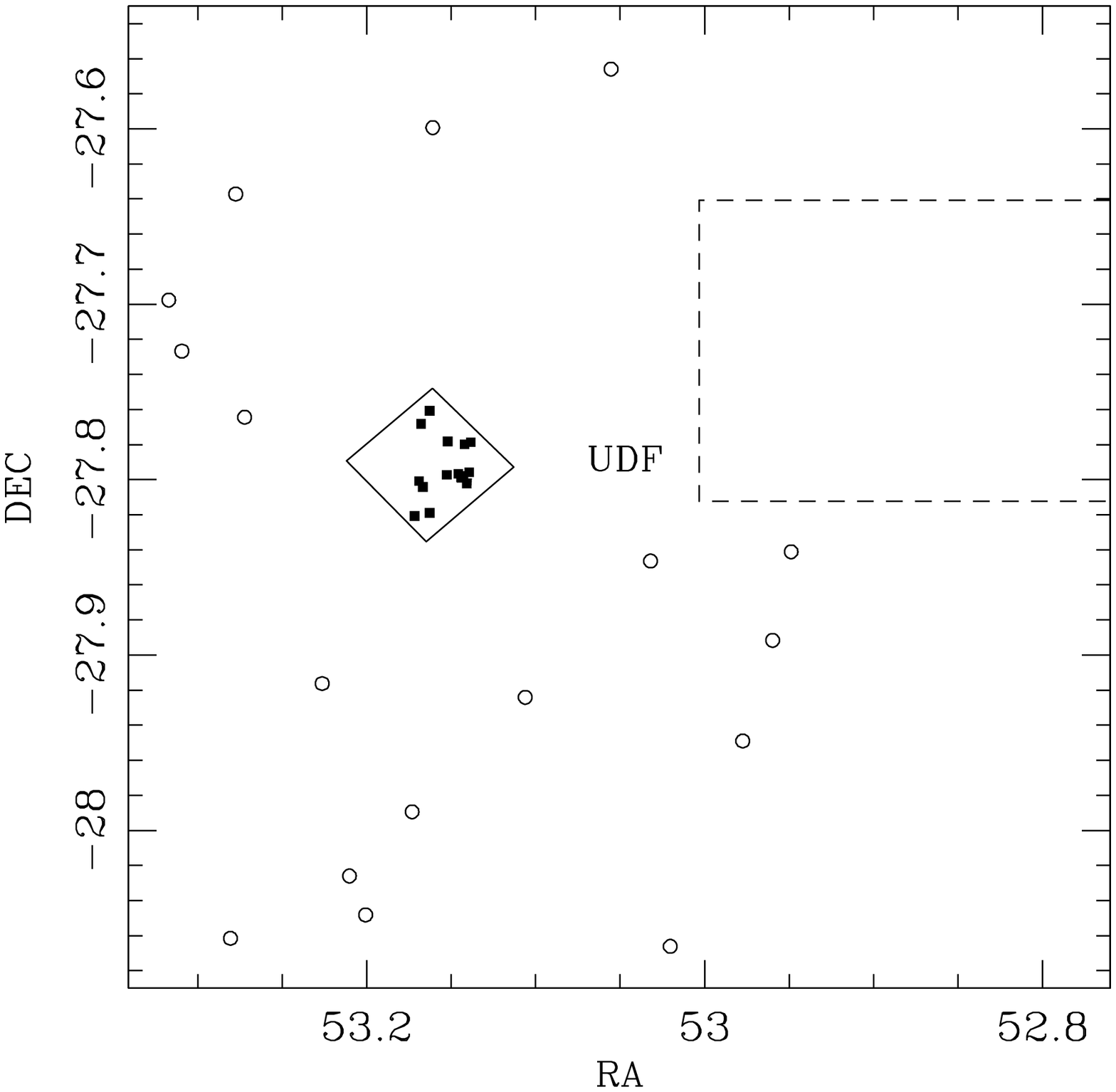}{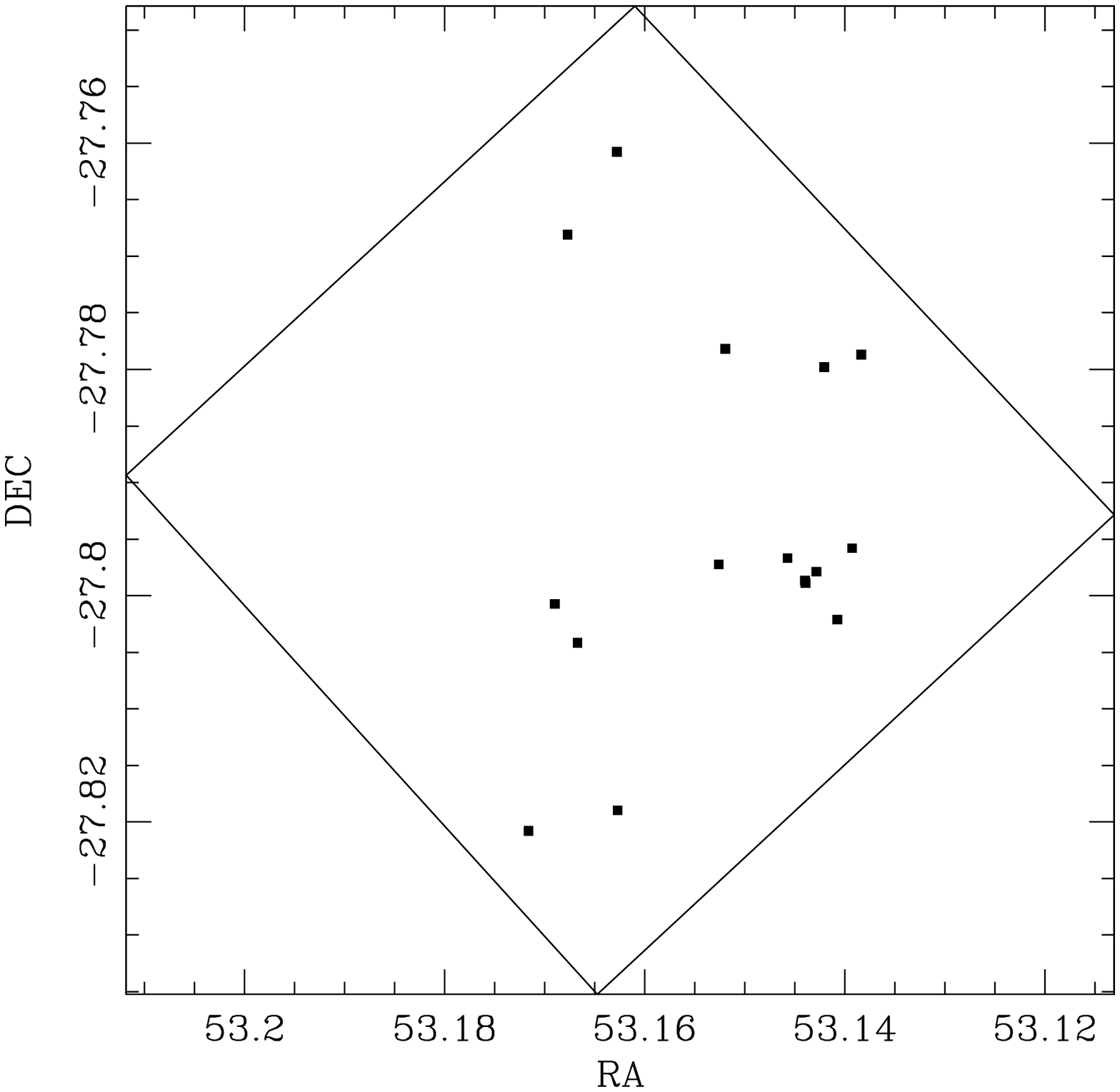}
\caption{(A) This figure shows the placement in the sky of the $z\sim 6$
galaxies in the UDF - seen here as square points in the diamond inset
which is the UDF coverage. The $z\sim 6$ galaxies avoid the eastern corner
of the UDF. A two-dimensional KS test shows that the probability of
this configuration by chance is about 5\%. A larger scale structure is
seen in the \lya\ emitters at z=5.7-5.77 which is consistent with the
UDF structure. The dotted rectangle shows the area not covered by the \lya\
survey due to a dead chip in the MOSAIC camera at CTIO.(B) shows the placement of 
galaxies at $z=5.9 \pm 0.2$ in the HUDF, (the left hand) half of the chip is 
devoid of galaxies. }
\label{lss}
\end{figure}

\subsection{Expectation of overdensity}

The location of the Hubble Ultra Deep Field was selected to include a
reasonably bright known galaxy at z=5.8 found in the GOODS survey
(Dickinson et al. 2004). Given that, we should consider what sort of
overdensity is expected. We calculate the overdensity in the galaxy
abundance in a cylindrical volume centered around a massive central
galaxy.

The first step is to compute the halo mass of the central galaxy.  The
total mass of a halo that corresponds to the abundance of 1 object
between $5.5 < z < 6.5$ in a solid angle of $\Delta\Omega=150$arcmin$^2$
is $M_c=9.1\times 10^{11}~{\rm M_\odot}$, which in turn corresponds to
a $3.7\sigma$ peak in the density field. This assumes a duty cycle of
$10^8$ yrs: each halo is assumed to be visible only a fraction
$f_d=10^8{\rm yr}/t_H(z=5.9)=0.11$ of the time, where $t_H(z)$ is the
age of the universe. The background cosmology is assumed to be flat
$\Lambda$CDM with $\Omega_{\rm m}=0.29$ and $H_0=72~{\rm
km~s^{-1}~Mpc^{-1}}$, consistent with the recent {\it WMAP}
measurements (Spergel et al. 2003).  The halo mass function was taken from
the fit to numerical simulations in Jenkins et al. 2001.

The second step is to estimate the typical expected galaxy overdensity
in the observed volume.  The volume is taken here to be a cylinder
centered at $z=5.9$ with angular diameter 3' (or $2R=7$ comoving Mpc
at $z=5.9$) and extending over the redshift range $z=5.9\pm 0.2$
(corresponding to a length of $2D=170$ comoving Mpc).

The mean overdensity $\langle\delta_g\rangle$ of galaxies of mass
$M_s$ expected to fall within this cylinder can then be found by
integrating the correlation function over the cylinder:

\begin{equation}
\langle\delta_g\rangle = 
\frac{1}{\pi(R^2-R^2_{\rm min})(D-R_{\rm min})} \times 
\int_{R_{\rm min}}^{R} 2\pi r dr
\int_{R_{\rm min}}^{D} dD
\left[1+b_c b_s g^2(z)\xi_m(x)\right].
\end{equation}

Here $g(z)$ is the linear growth function, $b_s$ and $b_c$ are the
linear halo bias \citep{mw96} of the ``satellite'' and central halos
with mass $M_s$ and $M_c$ respectively, $R_{\rm min}$ is a low-radius
cut, which we take to be the sum of the virial radii of the $M_s$ and
$M_c$ halos (in practice, the value does not matter), $\xi(r)$ is the
3D correlation function, and $x\equiv\sqrt{R^2+D^2}$ (the $\xi(r)$ subscripts
$g$ and $m$ refer to the values for galaxies and dark matter respectively.

We find $\langle\delta_g\rangle =1.78$ for $M_s=M_c=9.1\times
10^{11}~{\rm M_\odot}$.  This suggests that a factor of $\sim$two
overdensity should be typical in the observed volume. We note that
this overdensity is entirely a result of the high bias ($b=8.1$) of
the massive halos. As a result of the long redshift extent of the
overdense region (an order of magnitude larger than the correlation
length of $\sim 10$ Mpc), the corresponding mass overdensity
($1+b^2\langle\delta_m\rangle\approx 1+\langle\delta_g\rangle$) is
negligibly small, $\langle\delta_m\rangle =1.012$.  If we assume that the
redshift extent is reduced by a factor of 2 or 10 (changing $D$ from
85 Mpc to 42.5 or 8.5 Mpc), then we find $\langle\delta_g\rangle =2.5$
and 6.3, respectively. Thus higher resolution spectra may see a
more dramatic overdensity.

Three-dimensional effects can clearly be relevant. The above results
should motivate a study of the probability of intersecting the cosmic
filamentary density distribution in a way to reveal an elongated
overdense region with a large aspect ratio.

\section{Discussion and conclusions}

This paper establishes the fraction of color selected objects that are
{ \it bona fide\/} high redshift galaxies to be around 80\% for the color
$(i-z) > 0.9$ and magnitude limit of z=27.1--27.5. For a color selection
of $(i-z) > 1.3$ and magnitude limit of $ z' < 27.5$ mag, the success 
rate is 14/15, but such a color selection misses five blue
objects that are at redshifts $ z \approx 6$.

An overdensity is seen in the redshift distribution at $z=5.9 \pm
0.2$.  The galaxy number density in this redshift range is a factor of
two to four high. The extent of the redshift range spanned by this
overdensity may be smaller, because the grism spectra give redshifts
for LBGs accurate only to $\Delta z\approx 0.15$. \lya\ line emission and
absorption may further increase the observed scatter. Hence the
overdensity may be even larger in a small volume.  Hints of such an
overdensity were seen by Stanway et al.(2003), who found three \lya\
emitters at $z \approx 5.9$ in their spectroscopic followup. One of their 
objects is the same as object number 2225 in our sample. Further,
higher resolution spectroscopy would be invaluable in determining the
exact redshift extent and structure of this overdense region. A search
for \lya\ emitters has shown an overdensity at z=5.77 (823 nm)
relative to z=5.70 (815 nm) (Wang, Malhotra \& Rhoads 2005), as well
as a strong spatial gradient. The Hubble Ultra Deep Field sits at the
edge of this overdensity at z=5.77, and judging by the redshift
histograms seen in figure \ref{hist_z}, is part of it.  If we naively
multiply the spatial extent ($\approx 12'$) of the structure at z=5.77
seen by Wang et al.  2005 and the redshift extent ($\delta z = 0.4$)
of the overdensity seen in this paper, we would conclude that the
overdensity spans a volume of comoving $1.5 \times 10^5
Mpc^3$. According to current theories it would be hard to produce a
net overdensity of a factor of 4 over such a large volume, while an
overdensity of two could be due to the presence of a bright galaxy in
the UDF.

The overdensity at z=5.7-6.1, combined with a simple selection
function afforded by the grism, gives us an opportunity to derive the
luminosity function and star-formation rate in such an
overdensity. The complicating factor is the incompleteness in the
spectroscopic sample. As mentioned in section 2.1, we do not have
spectral information for 17 out of 46 objects brighter than \zfilt\ of
27.5. We can, however bracket the parameters of the luminosity
function by assuming that the spectroscopically confirmed sample
represents the lower limit and the upper limit is obtained by adding
to that all the objects for which we have no information. Figure
\ref{lumfn} shows the luminosity function for these two cases in the
redshift range $Z = 5.9 \pm 0.2$ and the best fitting Schecter
functions.  Following Yan and Windhorst (2004), we assume a slope of
$\alpha =-1.8$ and derive parameters $\mstar(z_{AB})=25.2$ and \phistar$=2.5
\times 10^{-4}$.  These values are comparable to the YW04 luminosity
function for $\alpha=-1.8$ which is $\mstar(z_{AB})=25.7$ and \phistar$=4
\times 10^{-4}$. While \phistar and \mstar have correlated error when fitting 
only upto \zfilt$=27.5$, the integrated
star-formation is fairly robustly estimated. The star-formation rate density (SFRD)
 derived from integrating over the luminosity function of the spectroscopically
confirmed sample is $2.5 \times 10^{-2} \msun Mpc^{-3} year^{-1}$, following the 
UV to SFR conversion in Madau, Pozetti \& Dickinson (1998).
Correcting for completeness with a factor of 1.6 gives $4 \times
10^{-2} \msun Mpc^{-3} year^{-1}$. All these values are significantly
higher than YW04 value of $1.2-1.5 \times 10^{-2} \msun
Mpc^{-3}year^{-1}$, consistent with there being an overdensity of at
least a factor of two in this redshift range. The lower bound to the
SFR is obtained by summing up the UV luminosity
in the objects spectroscopically confirmed and normalizing by the volume,
which is simply calculated as the comoving volume between $Z= 5.7-6.1$.
Thus the minimum SFR=$1.0 \times 10^{-2} \msun Mpc^{-3}year^{-1}$ which
is twice the estimate of Bunker et al. 2004 from the same field. Correcting
for spectroscopic incompleteness leads to SFR=$1.9 \times 10^{-2} 
\msun Mpc^{-3}year^{-1}$ which comes close to the required SFR 
needed for driving reionization especially if there are metal poor 
stars in these galaxies and the IGM
has a higher temperature (Stiavelli et al. 2004). The volume calculations 
above have taken the whole area of  the HUDF, whereas on the plane of the
sky $z \approx 6$ objects occupy about half (or less than) the area of the 
ACS chip (Figure \ref{lss}). Thus the volume over-density is a factor of 
four and the local SFR in the overdensity is definitely enough to drive
reionization of the local IGM.

By comparing with the luminosity function of \lya\ emitters, we see
that the space density normalization of LBGs in the UDF as measured
here and by Yan and Windhorst \phistar $ = 2.5 - 4 \times 10^{-4}$) is
two to four times higher than that of \lya\ emitters for which MR04
derive \phistar $ = 1 \times 10^{-4}$ at z=5.7. The \lya\ luminosity
function derived by MR04 is based on many surveys in different parts
of the sky, and therefore should be robust to cosmic variance. Given
the overdensity estimates of a factor of 2-4 here, the difference
between \phistar of the UDF LBGs and \lya\ galaxies is not
significant. The LBG space density could be consistent with that of
\lya\ emitters at z=5.7. That does not mean that every LBG has \lya\
emission, or that there is a one-to-one correspondence between the two.
That is because we often do not know the continuum luminosity of the
\lya\ galaxies, and they could be fainter than even this sample in
the rest UV.  On the flip side many of the LBGs in the present sample
could have weak \lya\ emission which would not be detected in low
resolution spectra.

The SFRD derived from \lya emitters is
$1.8-3.6 \times 10^{-3}$, which is roughly a tenth of the SFRD derived
here. This is consistent with the fact that SFR for individual \lya\
galaxies, derived from the \lya\ line alone, is on the average 1/10th
that of a typical LBG (Rhoads et al. 2000, 2003, Dawson et al. 2004). 

\begin{figure}[ht]
\epsscale{0.5}
\plotone{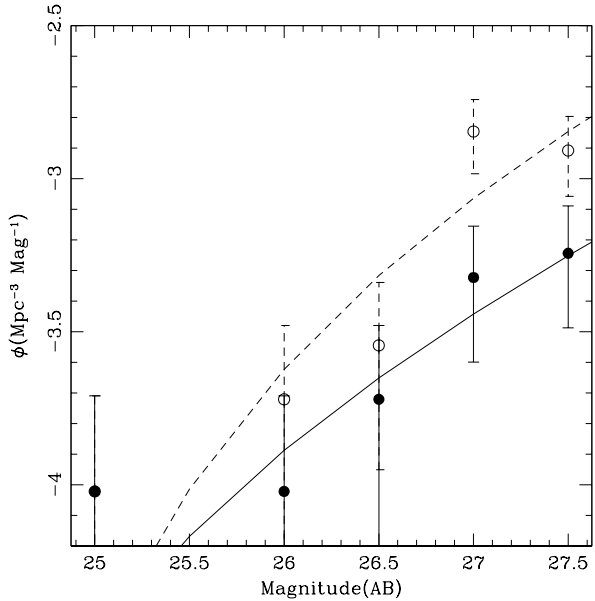}
\caption{The solid points show the luminosity function of the 
spectroscopically confirmed galaxies in the redshift range z=5.7--6.1
where an overdensity is seen; and the solid line as the best fit
Schecter function, assuming a faint end slope of $\alpha = -1.8$)
giving the best fit values of (\phistar, \lstar)=(25.2, 2.5$\times
10^{-4}$). The open circles show the data corrected for incompleteness
of the spectroscopic sample, by adding the unidentified objects to the
appropriate magnitude bin. The resulting parameters are (\mstar,
\phistar, $\alpha$)=(25.9, 4$\times 10^{-4}$,-1.8)} \label{lumfn}
\end{figure}

This sample of spectroscopically confirmed Lyman break galaxies can
also provide an independent test of reionization.  If the IGM is
neutral at $z > 6$, the fraction of Lyman break galaxies that show
\lya\ line emission should drop, because the damping wings of neutral
IGM reduce the line to at most 1/3rd its original strength (Haiman
2002, Santos 2004). Prior to this paper, all the spectroscopically
confirmed galaxies at $z > 6$, and all but two at $z > 5$, showed
\lya\ emission. This may be a selection effect, since objects with
strong lines are easier to confirm spectrocopically. For reference,
only 25\% of LBGs show \lya\ emission at $z \simeq 3$ (Steidel et
al. 1999).  Now we have a sample of about 23 Lyman break galaxies
(LBGs) selected using low resolution spectra from the grism on HST at
redshifts $z=5.4$--$7$, not biased by the presence or absence of \lya\
line.  Deep, higher-resolution spectra of these objects should be able
to detect lines with rest-frame EW of 30 \AA.  This would determine
the \lya\ emitter fraction and thus constrain whether the IGM is
neutral.

\acknowledgements 

We are grateful to the STScI director Steve Beckwith and the Hubble
Ultra Deep Field team for observing and making available the excellent
imaging data. Thanks are also due to Rodger Thompson and his team for
making available the NICMOS observations of the same region.  We thank
the referee for a prompt and helpful report which improved this
paper. The imaging and spectroscopy data are based on observations
with the NASA/ESA {\it Hubble Space Telescope}, obtained at the Space
Telescope Science Institute, which is operated by AURA Inc., under
NASA contract NAS 5-26555. This work was supported by grant
GO-09793.01-A, GO-09793.02-A, GO-09793.03-A and GO-09793.08-A from the
Space Telescope Science Institute, which is operated by AURA under
NASA contract NAS5-26555. ZH was supported in part by NSF through
grants AST-0307200 and AST-0307291 and by NASA through grant
NAG5-26029. This project has made use of the aXe extraction software,
produced by ST-ECF, Garching, Germany.

\appendix

\section{Grism Spectra} 
In this appendix we present the grism spectra of 23 high redshift
objects. On each, we superpose the best fit Lyman break spectrum ,
represented by a power law of $F_{\lambda} \propto \lambda
^ {2.2}$ and Madau IGM absorption (Madau 1995) (top panel). We also
see a consistency with the broad-band fluxes (middle panel). The
agreement with the broad band fluxes is not perfect due to aperture
mismatches, which we have tried to minimize and because the broad-band
fluxes were not used in fitting, and the UV slope of the galaxies does
vary from object to object, unlike the models we have used. The quality
of the data, the s/n seen in most sources, and the limited wavelength
range redward of the Lyman break simply did not support 
fitting the slope as an extra parameter.

The error-bars on the near-infrared fluxes were estimated by placing
random apertures of a range of sizes on the finished image and
estimating the noise properties of those apertures. The error-bars 
calculated this way are larger than the typical errors quoted elsewhere
(Bouwens et al. 2004), but reflect a more realistic picture. A similar
exercise with ACS images gave error estimates that are similar to the
formal error estimates obtained by multiplying the rms deviations with
the square-root of the number of pixels.

\begin{figure}
\epsscale{0.7}
\plotone{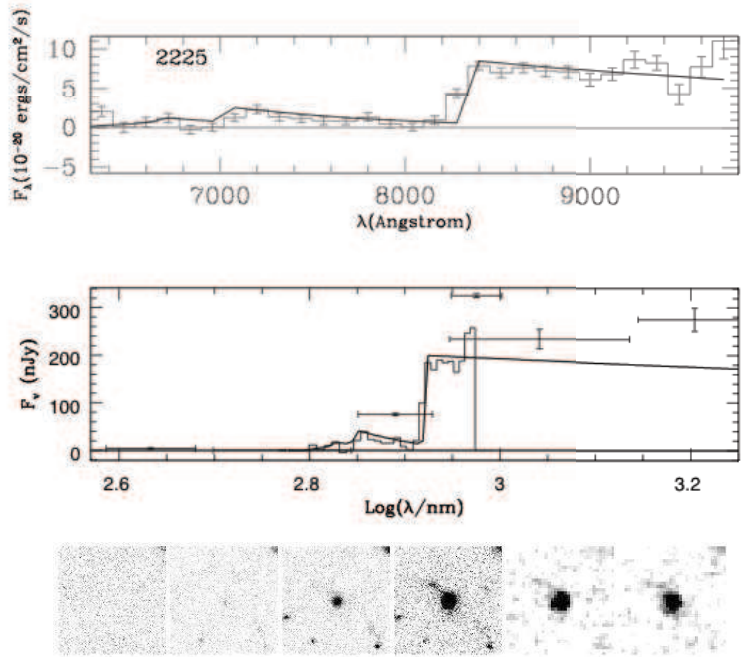}
\vspace{0.9in}
\caption{
In this figure we present the grism spectra and the best fit using IGM
absorption; the top panel show the fits in $F_\lambda$ vs. wavelength.
 The middle panel show the same objects in $F_\nu$ vs.  log of
 wavelength, showing also the consistency with broad-band colors.  The
 bottom panel shows cutout images ( 3$\arcsec$ on the side) in B,V,
 \ifilt, \zfilt, F110W and F160W. The stretch is each image is
 adjusted to go from -3$\sigma$ to 5$\sigma$. The top panel is labeled with 
the object name {\bf Due to size restrictions
 not all the spectra could be accomodated in the astro-ph version.
 Please go to www.stsci.edu/$\sim$san/spectra.pdf for a supplement
 that shows all the spectra}}
\label{spectra}
\end{figure}

\clearpage
\begin{figure}
\epsscale{1.0}
\plotone{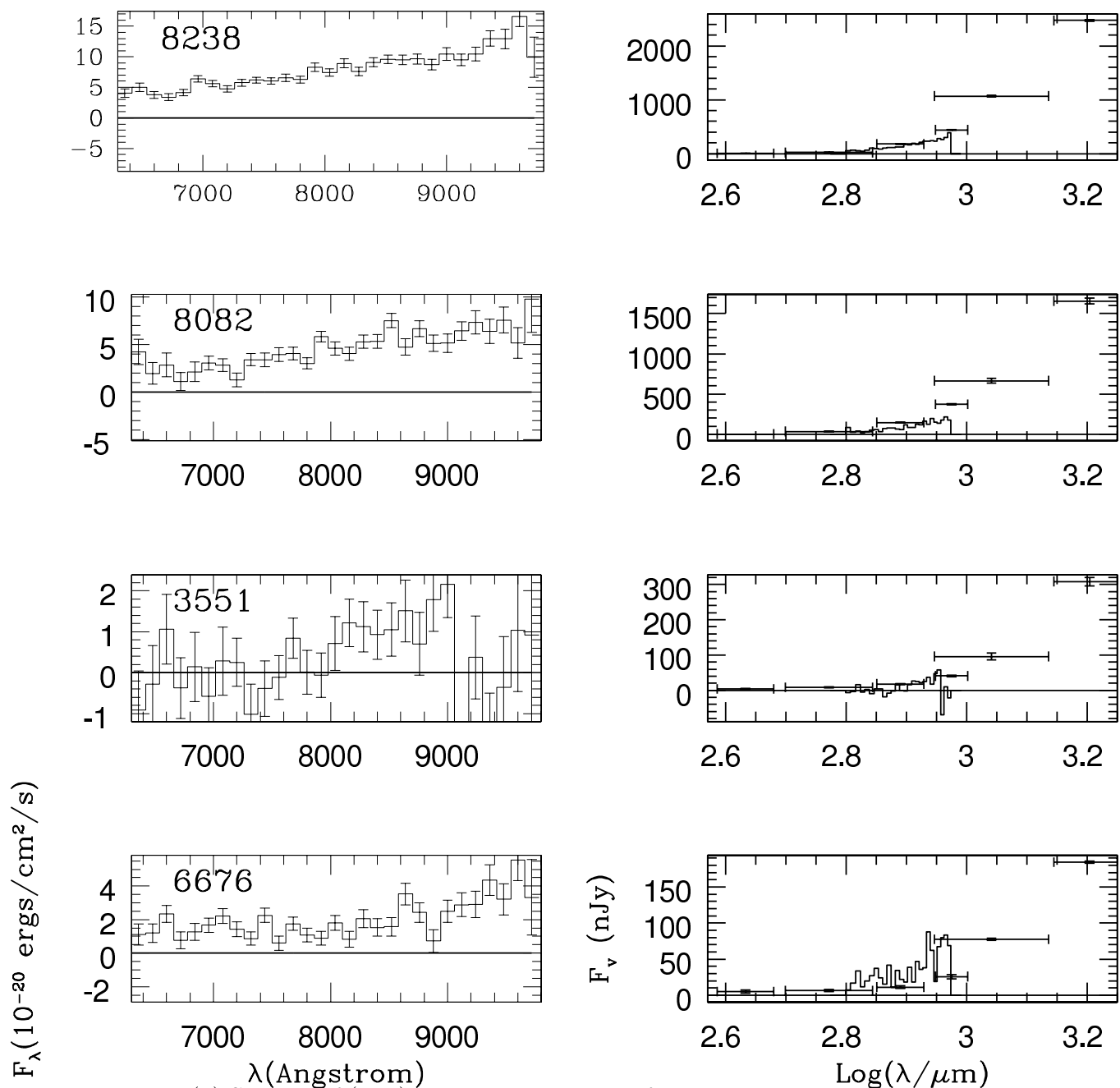}
\caption{ (a) Spectra of (i'-z') selected red objects, for which the spectra
 do not show a break  but a steady rise of flux into the red. Some of these 
are  old, elliptical galaxies ( UDF~8238 is further discussed by Daddi et
al. 2005); and some may be dusty, star-forming galaxies}
 \label{spectra-red} 
\end{figure}

\begin{figure}
\epsscale{1.0}
\plotone{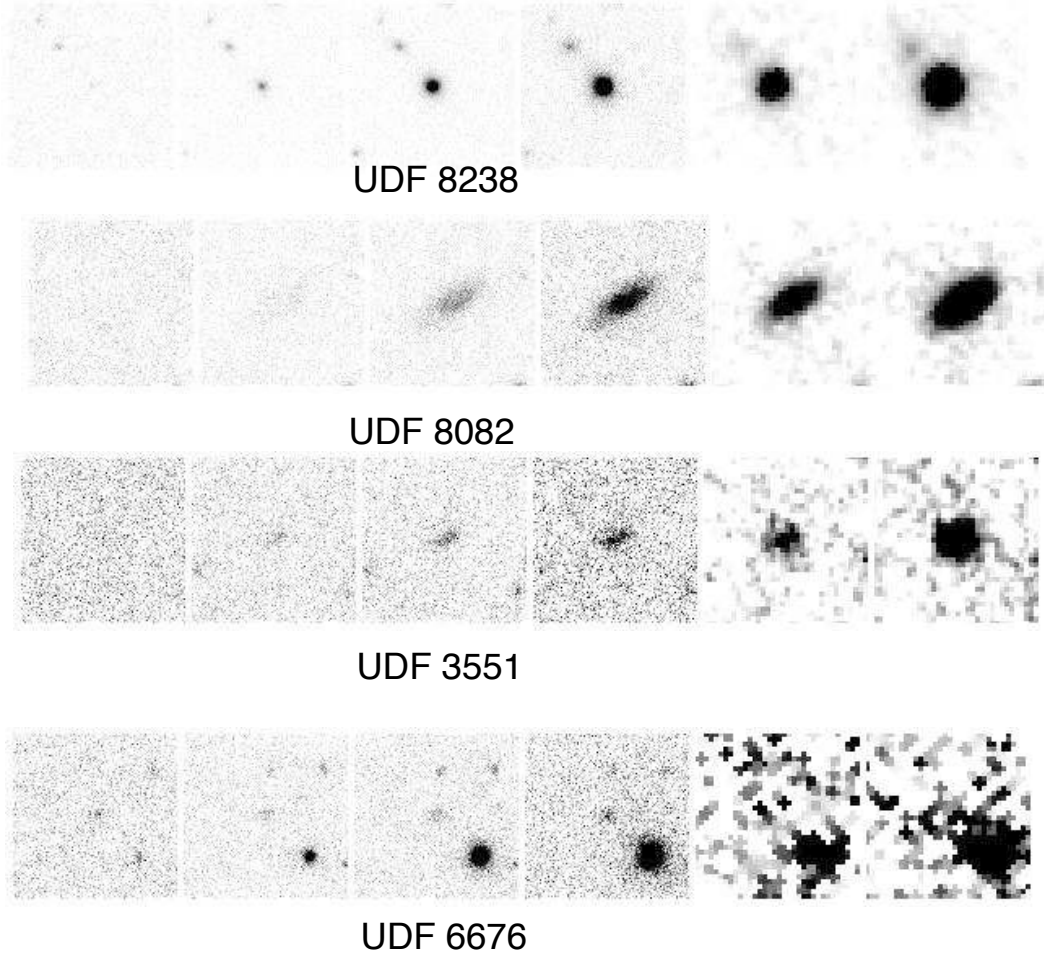}
\vspace{-0.4in}
\caption{ (b) Multiband images of color 
(\ifilt - \zfilt)$ > 0.9$ selected
red objects, for which the spectra do not show a break but a steady
rise of flux into the red. Some of these are old, elliptical galaxies
( UDF~8238 is further discussed by Daddi et al. 2005); and some may be
dusty, star-forming galaxies. The images are $3" \times 3"$ on the side and
are (going left to right) in the bands B,V,\ifilt,\zfilt, F110W( $\approx$ 
J band) and F160W ($\approx$ H band)}
 \label{images-red} 
\end{figure}

\clearpage

\begin{deluxetable}{ccccccc}
\tablecolumns{7}
\tablewidth{0pc}
\tablecaption{Spectroscopic redshifts of i-droputs }
\tablehead{
\colhead{UDF id} &
\colhead{\zfilt}  &
\colhead{\ifilt-\zfilt} &
\colhead{Redshift} &
\colhead{RA (J2000)} &
\colhead{DEC(J2000)} &
\colhead{S/N}}
\startdata
  2225 \tablenotemark{a} & 25.06 & 1.55 &  5.8  & 03:32:40.012 & -27:48:14.97 &  23.4 \\
  9202       &  27.43    &   1.56 &5.7  & 03:32:33.207 & -27:46:43.26 &  6.4\\
  2690       & 27.33   &    1.7 &  5.9  & 03:32:33.781 & -27:48: 7.59 &  7.0 \\
  32521     & 26.82   &    2.4 &   5.9 & 03:32:36.626 & -27:47:50.06 &  6.7 \\ 
  3377/3398 \tablenotemark{b}  & 26.49  &  1.0,1.3 & 5.6  & 03:32:32.636 & -27:47:54.30 &   6.2   \\
  9857       &  26.95  &     1.4 & 5.8  & 03:32:39.066 & -27:45:38.75 & 6.3  \\ 
  6329       &  26.88   &    1.1 & 5.5  & 03:32:35.196 & -27:47:10.08 &   5.4   \\
  8961       &  26.54   &    2.12 & 5.8  & 03:32:34.097 & -27:46:47.23 &   5.3 \\
  8033       &   26.05  &     1.9 &6.0  & 03:32:36.467 & -27:46:41.44 &   5.2   \\	
  32042      &  28.2  &    2.4 &5.75  & 03:32:40.554 & -27:48: 2.61 & 4.9 \\
  36383     &  28.0  &     3.0 &5.8  & 03:32:40.249 & -27:46: 5.18 & 4.8 \\
  457       &  28.0 & 1.0 & 5.8  & 03:32:39.048 & -27:49: 8.30 &  4.6 \\
  4050      &  27.33 &      1.9 & 6    & 03:32:33.429 & -27:47:44.86 &   4.5  \\
  322        &  26.91 &      1.9 &5.7  & 03:32:41.187 & -27:49:14.85 & 3.9 \\
  3317       &  26.94  &     1.15 & 6.1  & 03:32:34.556 & -27:47:55.15 & 4.3 \\	
  3503       & 27.7 & 1.7 & 6.4  & 03:32:34.306 & -27:47:53.54 & 4.3 \\
  3807	   &  27.73 &  0.97 & 6.1  & 03:32:34.976 & -27:47:48.05 & 3.9 \\
  3325       &  27  &   1.85 & 6.0 & 03:32:34.547 & -27:47:55.98 & 3.7\\  
  3450      &  27.05 &     1.5 & 5.9  & 03:32:34.283 & -27:47:52.35 &  3.4 \\
  35506      & 27.46  &     2.8 & 6.15 & 03:32:39.860 & -27:46:19.08  &  3.1 \\
  33003     &   27.8    &   3.3 & 6.4  & 03:32:35.056 & -27:47:40.18 & 3.3 \\
  30591     &  27.13    &   2.56 & 6.7  & 03:32:37.277 & -27:48:54.57 & 3.5 \\
\multicolumn{7}{l}{Marginal detections} \\
  2631      &  27.71   &    2.262 & 6.6  & 03:32:42.596 & -27:48: 8.83 & 2.9 \\ 
\multicolumn{7}{l}{Red galaxies at intermediate redshifts} \\
\multicolumn{7}{l}{8238, 8082, 3551, 1238, 6676}  \\
\multicolumn{7}{l}{Dwarf stars}  \\
\multicolumn{7}{l}{443, 366} \\
\enddata
\tablenotetext{a}{UDF 2225 has spectroscopic confirmation with Keck(z=5.83, Dickinson et al. 2004), and Gemini (z=5.83), Stanway et al. 2004}
\tablenotetext{b}{These two objects are identified as seperate objects 
in the UDF catalog, but lie close to each other and show similar spectra, thus we consider them to be the same object}
\end{deluxetable}

\end{document}